\documentclass[aps,superscriptaddress,showpacs,nofootinbib]{revtex4}

\usepackage{epsfig,rotating}
\input{epsf}

\begin{document}

\title{The warm inflationary universe}

\author{Arjun Berera}
\affiliation{ School of Physics, University of
Edinburgh, Edinburgh, EH9 3JZ, United Kingdom}

\begin{abstract}

In the past decade, the importance of dissipation and
fluctuation to inflationary dynamics has been realized
and has led to a new picture of inflation called warm
inflation.  Although these phenomena are common to
condensed matter systems, for inflation models
their importance has only recently started to be appreciated.
The article describes the motivation for these phenomenon during
inflation and then examines their origins from first principles
quantum field theory treatments of inflation models.
Cosmology today is a data intensive field and this is driving
theory to greater precision and predictability.
This opens the possibility to consider 
tests for detecting observational signatures of dissipative
processes, which will be discussed.   In addition it
will be discussed how
particle physics and cosmology are now working in tandem 
to push the boundaries of our knowledge about fundamental physics.  

\medskip

\medskip

\end{abstract}

\pacs{98.80.Cq, 05.70.Ln, 11.10.Wx}

\maketitle

\medskip

Published in Contemporary Physics {\bf 47}, 33 (2006).

\section{Introduction}
\label{intro}

By the 70s the hot Big Bang model had proven very successful
in explaining both general qualitative features of our Universe,
such as its expansion and presence of a background radiation,
as well as providing a quantitatively successful theory of nucleosynthesis
of the light elements.
However there were difficult problems within this model, which in
the early 80's were shown could be nicely solved through
the inflation idea.  Since its introduction, inflation has
been a theoretically very compelling picture in search
of justification from observation.
In the past decade, precise data on the cosmic microwave
background radiation (CMB) obtained primarily 
from satellite experiments has provided support for inflation.
This success has summoned growing interest both in constructing
realistic particle physics motivated models of inflation
and in calculating predictions from these models as accurately
as possible.   One outcome that has emerged from examining inflation
models at close scrutiny has been that dissipative and nonequilibrium
effects during the inflation phase
are prevalent in many such models.  This has led to an understanding
that there are two distinct dynamical realizations of inflation,
which have been termed cold and warm inflation.  Cold inflation
coincides with the original inflation picture and has been
the subject of many reviews \cite{cireview}.  The purpose of this paper
is to explain warm inflation and review progress in developing
this idea.

Warm inflation is not a radical rethink of inflation, but rather
a completion of the original picture.  The idea stems from
an elementary observation.  
The central theme of
inflationary dynamics today is the evolution
of a scalar field, which during inflation carries most of the energy
of the universe and which interacts with other fields.
On the one hand, the standard inflation picture 
makes a tacit assumption that these
interactions have no other effect except to modify the scalar
field effective potential through quantum corrections.
On the other hand, the warm inflation picture 
is that interactions not only
modify the scalar field potential but also lead
to fluctuation and dissipation effects.
Certainly in condensed matter systems the effect of interactions
generally lead to all three of these effects
(several examples are given in \cite{weiss}).
Also from a statistical mechanics perspective
the scalar field would want to dissipate its energy
to other fields and the whole system would try
to equally distribute the available energy.
Ultimately the question must be answered through a thorough dynamical
calculation.  

Dynamics in cosmology differs in one important
way from condensed matter systems, in that all processes occur
in the presence of an expanding universe.
Expansion acts like an external force
which is constantly altering the state of any cosmological
scale system.  For example energy in the universe is continually
being diluted due to expansion.  Similarly, the configuration
of any cosmological scale process is being altered over time.
Thus if the quantum mechanical processes that lead to dissipation
operate at a time scale much slower than the expansion rate
of the universe, than these processes could be totally shut down
due to expansion, even if the same processes in a nonexpanding system, like
a condensed matter system, operates efficiently.
This is the important question that must be understood.

The problem being posed
here may appear very specialized and technical.  Quite
oppositely, what makes the problem interesting is that
it combines simple physics from very different
disciplines, i.e.  cosmology, particle physics and condensed
matter physics. The challenge is
in appreciating how these different ideas join together,
and that is what is to be explained.
The other reason this problem is interesting,
which also will be discussed in this article,
is that the consequences
of dissipation and fluctuation during inflation have distinct
model building implications for particle physics that do not arise
in cold inflationary dynamics, where these effects
are negligible.
This article is fully self-contained, starting from
the relevant background in General Relativity and inflation
before addressing the main topic.
For the reader familiar with these preliminaries,
they can skip the next four sections and start immediately
from Sect. \ref{sect6}.

\section{Basic equations of cosmology}
\label{sect2}

The single most significant fact about our universe is that all
galaxies, except a few nearby ones, are receding from us.  Moreover,
the redshift, a measure of the recession speed, is larger for
the fainter more distant galaxies.  This suggests that our universe
is expanding, with clusters of galaxies getting more widely separated,
and so more thinly distributed through space, as time goes on.

To understand this expansion, or any other aspect of the dynamics
of the universe, Einstein's General Theory of Relativity must be used.
In General Relativity, matter and geometry are intrinsically
entangled.  General Relativity treats gravity as a deformation
in the geometry of space and in the flow of time, due to
the presence of matter.  An analogy would be to
represent space by a stretched membrane.  For a very light ball
rolling on it, the surface would be flat.
Now imagine placing a heavy ball on this membrane.  Its effect would
be to depress the membrane in its vicinity.  
The light ball now would no longer roll in a straight line, 
but would curve in the direction of the heavy ball.
The depression in the membrane is analogous to gravity
which is produced due to the presence of matter, in
this case the heavy ball.  So just as in Einstein's picture,
the acceleration of the light ball due to the presence of gravity
is viewed as arising due to a deformation in the geometry of space.

The geometry of the universe is expressed through the metric
tensor $g_{\mu \nu}$ whereas the matter is described through 
the energy-momentum tensor $T_{\mu \nu}$.
The geometry and matter are related through the Einstein equation
\begin{equation}
R_{\mu \nu} - \frac{1}{2} g_{\mu \nu} R = -8 \pi G T_{\mu \nu},
\label{ee}
\end{equation}
where $G$ is the Newton constant of gravitation,
$R_{\mu \nu}$ is the Ricci tensor,
and $R \equiv g_{\mu \nu} R^{\mu \nu}$ is the curvature.
The Ricci tensor is constructed from various derivatives of
the metric tensor $g_{\mu \nu}$ and gives a measure of
the curvature of space-time.
The Einstein equation (\ref{ee}), as can be seen, is an equation relating
tensors, here four by four matrices, and so 
in general there are 16 separate
equations.  All gravitational phenomenon are governed by these
equations.  

Our interest is cosmology, meaning
phenomena which in each epoch
occurs on time-scales of order the age of the universe at that time
and on distance scales as large as the size of the universe at
the given time.
On these cosmological scales, the foundations of the whole
subject is embodied in the cosmological principle, which states
no observer occupies a preferred position in the universe.
This principle implies the universe must be homogeneous
(looks the same from every point) and isotropic (looks the same
in all directions).  Note that homogeniety does not automatically
imply isotropy, with a simple counterexample being a universe
with a uniform magnetic field; in such a case, every point
looks the same as every other, but there is a preferred direction.
However isotropy about every point does imply homogeniety,
giving a universe in which matter is smoothly distributed.

To quantify the cosmological principle,
in regards the geometry, the metric of a space-time which is
spatially homogeneous and isotropic is given by the
Friedmann-Robertson-Walker (FRW) metric,
\begin{equation}
ds^2 = dt^2 - a^2(t) \left[\frac{dr^2}{1-kr^2} + r^2 d\theta^2 +
r^2 \sin^2 \theta d\beta^2 \right],
\label{frw}
\end{equation}
where $(r,\theta,\beta)$ are spherical-polar coordinates parametrizing
the spatial dimensions, $t$ is cosmic time, and $k=1,-1,0$ describe
spaces of constant positive, negative and zero spatial  curvature.
The most important quantity here is $a(t)$ which is the cosmic
scale factor and describes the expansion of the universe.

For the metric Eq. (\ref{frw}), the Einstein
equations  Eq. (\ref{ee}) reduce into two independent equations,
one the celebrated Friedmann Equation
\begin{equation}
\frac{{\dot a}^2}{a^2} + \frac{k}{a^2} = \frac{8 \pi G}{3} \rho,
\label{fe}
\end{equation}
and the other an equation no one ever bothered to name 
\begin{equation}
2 \frac{\ddot a}{a} + \frac{{\dot a}^2}{a^2} + \frac{k}{a^2}
= - 8 \pi G p .
\label{unke}
\end{equation}
The expansion rate of the universe is determined by the
Hubble parameter $H \equiv {\dot a}/a$, which is the single
most important parameter characterizing the universe.
The Hubble time $H^{-1}$ sets the time scale for expansion,
with the scale factor $a$ doubling in approximately
one Hubble time.  This fact also means the Hubble length
$c/H$ sets the scale of the observable universe, meaning
the distance a light beam has traversed during the age of
the universe.  The Hubble parameter today $H_0$, relates the
velocity $v$ of the distant galaxies receding away from us
to their distance, $r_0$,
from us via the Hubble law $v = H_0 r_0$.

The matter content in the FRW model is described through just two
quantities, the energy $\rho$ and pressure $p$ densities, which in
the simplest models are related through the equation of
state $p = w \rho$.
Given an equation of state $w$, one can then solve the cosmological
Einstein equations (\ref{fe}) and (\ref{unke}).
Three cases are of particular importance for describing the cosmology
of our universe
\begin{eqnarray}
{\rm radiation} \ \ p & = & \frac{1}{3} \rho \ \ \Rightarrow \rho \propto a^{-4}, \ \ a \propto t^{1/2},
\nonumber \\
{\rm matter} \ \ p & = & 0 \ \ \Rightarrow \rho \propto a^{-3}, \ \ a \propto t^{2/3},
\nonumber \\
{\rm vacuum} \ \ p & = & - \rho  \ \ \Rightarrow \rho \propto const., \ \ \ln(a) \propto t .
\label{bbsol}
\end{eqnarray}
One comment here on terminology.
The vacuum energy is also often referred to as the cosmological
constant, with particle physicists preferring the former 
and General Relativists the latter.

\section{Standard cosmology}
\label{sect3}

What is called the Hot Big Bang model, and also referred to
as the Standard Cosmological Model, is characterized by a universe in
which large scale gravitational attraction causes the universe
to expanding at a decelerating rate ${\ddot a} < 0$.
In particular, the Big Bang model is comprised of
the first two solutions from Eq. (\ref{bbsol}), along with some
general knowledge about the particle content of the universe
and some basic statistical mechanics.  Based on this information,
the Big Bang model asserts that the late phases of the
universe right up to the present are in a matter dominated era, which means
a period when the evolution of the scale factor goes as
the matter case in Eq. (\ref{bbsol}) (there is
evidence that in the very recent epoch the universe might
be in a regime dominated by vacuum energy.  This detail
is not relevant to our discussion and so we will
not delve into it.  There are several reviews on this 
subject \cite{coscon,coscon2}).  
The expansion of
the universe implies the earlier universe gets increasingly denser and 
hotter.  Going back in time, the Big Bang model finds 
that at some stage in the past the universe was in a radiation
dominated regime which characterizes the early phase of the universe. 

Since the expansion implies the younger universe was smaller and hotter,
based on our current knowledge of physics, this leads to a picture
of the universe when it was only a few minutes old at at a temperature
of $10^{10} K$.  The significant point is this picture can be
tested.  One of the most important tests is that the
light elements are predicted to be `cooked' at this time,
a process called Big Bang nucleosynthesis (BBN).
Thus we can compare calculations of the cosmological fraction
of elements like hydrogen, helium, deuterium and lithium with
their observed abundances.  For example the visible matter
content in the universe is found to be about 23$\%$ helium,
and this is in excellent agreement with predictions from
the Big Bang model.  This result is a more stringent verification
of the Big Bang model than what might meet the eye.  For example
it may have happened that astronomers had found astrophysical
scaled objects whose helium abundance was far below this prediction
of the Big Bang model.  This would have been fatal, since
extra helium made in stars could boost helium above its pregalatic
abundance, but there appears to be no way of converting all helium back
into hydrogen.  Another verification of the Big Bang
model, from nucleosythesis in the early universe is
deuterium (heavy hydrogen) abundance.  There is only a tiny trace
of deuterium in the universe, its abundance ratio to hydrogen is
about $1/50,000$ and this is consistent with predictions
from the Big Bang model.

Another robust prediction of the Big Bang model is the CMB.
In the past decade the CMB has been very accurately measured 
in particular by two satellite experiments,
the Cosmic Background Explorer (COBE) and
the Wilkinson Microwave Anisotropy Probe (WMAP)
(for reviews see \cite{al,hp}).
These experiments have found the CMB to have a nearly perfect
blackbody spectrum at temperature $T_0=2.725$K.
This result is another stringent test of the Big Bang model.
For example it may have happened that the observed CMB
spectrum differed too drastically from the expected blackbody
form.  Also, there are tiny fluctuations in the CMB temperature
distribution at one part in $10^5$.
The Big Bang model predicts certain 
correlation between fluctuations in the CMB spectrum and matter
fluctuations in the universe.
It could have happened,
that in fact these correlations were not found.
The failure of either of these two tests could have
been a major blow to the Big Bang model.
However every time the data has shown consistency with the model.

\section{Cosmological puzzles}
\label{sect4}

In short, the Big Bang model is amazingly successful.
It provides a testable and consistent account of the early universe,
which has been verified at least since the era of Big Bang
nucleosynthesis ($t \sim 1 {\rm sec}$) up to the present.
Most importantly there is no observational nor
experimental data that casts any serious doubts on this model.
Nevertheless the model does not explain everything.
There are some serious shortcomings to this model, which
are often referred to as the cosmological puzzles.
These puzzles are not inconsistencies within the Big Bang model,
but rather they are questions that the model itself allows
to be asked, but within its limits can not answer.

\begin{figure}[ht]
\vspace{1cm}
\epsfysize=5.95cm
{\centerline{\epsfbox{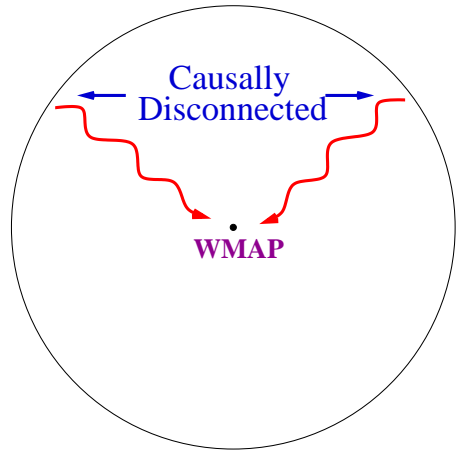}}}
\caption{CMB photons observed from very different directions in the sky
would never have been in causal contact according to the Big Bang model}
\label{fig1}
\end{figure}

The first and foremost puzzle is that the universe is homogeneous
on a much bigger scale than the Big Bang model would
predict.  This fact is seen most evidently in the CMB.
The CMB radiation that we receive from all directions in the
sky was emitted about $10^6$ years after the birth of the universe.
Based on this fact, it would mean for example that the CMB photons
in Fig. \ref{fig1}
received today from opposite directions in the sky
were separated by about $10^7$ light years at time of emission.
Thus they could not have had any causal connection with each other.
In this case how did they know to be at 2.725K temperature within 
parts in $10^5$?  This is often called the horizon problem.

Closely related to the horizon problem is the formation
of structure problem.  Observations indicate nonrandom correlations
in the distribution of galaxies
and clusters of galaxies on scales much larger than $50 \ {\rm Mpc}$.
Beyond this scale, it is very difficult to see how such correlations
could have been created in Standard Cosmology.
Similarly, the CMB spectrum of fluctuations 
at time of last scattering is found to be almost
scale invariant up to the largest observable scales in the universe
$\sim 3000 {\rm Mpc}$.  In the Standard Cosmology, no causal
mechanism can explain this.

The other major puzzle has to do with the global geometry of
the universe.  Below Eq. (\ref{frw}) we briefly touched on
the fact that in the FRW model, the universe can have a global
geometry that is either open, flat or closed.  What does that mean?  
If the universe is flat. than it is described by Euclidean geometry, so
two parallel lines would never cross.  However the universe may
also be open, in which case parallel lines diverge from one another,
or it may be closed, like the way the surface of a sphere is
closed, so that parallel lines cross one another like
the meridian lines on a globe.  What is found from observation
is that our universe today is very close to being flat.
It has a radius of curvature of order the scale of the observed
universe $\sim 10^{28} {\rm cm}$.

Based on General Relativity and
the Big Bang model, one can ask, what sort of global geometry should
we expect.  In GR the only natural length scale is the Planck Length,
$l_p \sim 10^{-33} {\rm cm}$.  Thus one might expect the typical radius
of curvature to be of order this scale, but that falls
some 60 orders of magnitude below observation.
A more useful approach is to solve the cosmological
Einstein equations (\ref{fe}) and (\ref{unke}), which should tell us given an
initial condition at some early stage of the universe, what
the geometry of the universe is today.  What one finds is that
in the Big Bang model, for either matter of radiation dominated
cases, the flat geometry is a repulsive point.  This means
whatever initial conditions one chooses, whether for a geometry that
is open of closed, as time evolves, the universe gets further and
further away from the flat geometry, i.e. a open universe becomes
more open and a closed one more closed.  So the only way
the Big Bang model could explain the present
near flatness of our universe is if in its early phases the
universe was tuned to be very very flat.
Typically in physics whenever an extreme tuning is necessary,
the usual assumption is that some aspect of the problem has
been understood incompletely.
This problem is often referred to as the flatness problem.

As mentioned in the beginning of this section, one now can see
that none of the problems mentioned above imply the Big Bang
model is flawed, but rather that it is potentially incomplete.
One could simply argue that the initial conditions of the
universe were such that the temperature happened
to be the same in all the causally
disconnected regimes we observe today 
and the universe in its early phases just happened to be
very very close to flat.  The problem is these are
fine tuning argument.  In such a circumstance,
one is compelled to ask whether there is some dynamical
mechanism that has not been appreciate which 
provides a simple explanation for
these peculiar features.

\begin{figure}[ht]
\vspace{1cm}
\epsfysize=5.95cm
{\centerline{\epsfbox{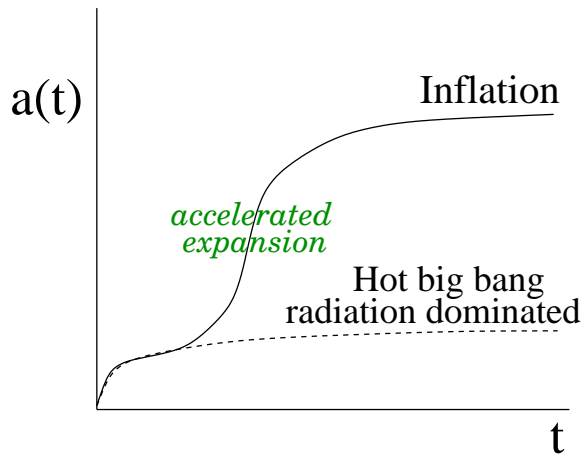}}}
\caption{A schematic representation of how the 
scale factor in the inflation regime grows much faster
than in the Big Bang regime}
\label{fig2}
\end{figure}

\section{Kinematics of inflation}
\label{sect5}

As it turns out, one simple and elegant idea, called inflation,
can solve both these cosmological puzzles and in a way
that nicely fits together with the Big Bang model.
The most general description of inflation is that it is
a phase in which the scale factor grows at an accelerating rate
${\ddot a} > 0$. 
To study inflation, it is often convenient to use an
equivalent form of the cosmological Einstein equations (\ref{fe})
and (\ref{unke}), the scale factor equation
\begin{equation}
\frac{\ddot a}{a} = - \frac{4 \pi G}{3}(\rho + 3p)
\label{scalee}
\end{equation}
and the energy conservation equation 
\begin{equation}
{\dot \rho} = -3 H (\rho + p)
\label{econs}
\end{equation}
The scale factor equation is obtained by subtracting
Eq. (\ref{unke}) from Eq. (\ref{fe}) and the energy conservation
equation is obtained by taking the time derivative of
Eq, (\ref{fe}) and then use Eq. (\ref{unke}) to eliminate certain terms.

From the scale factor equation it can be seen that in order
to obtain an accelerating scale factor ${\ddot a} > 0$,
it requires $p < - \rho/3$, thus a substance with
negative pressure.
This means a universe where the dominant form
of matter produces a repulsive form of gravity.
The most common example of such a scale factor behavior is exponential,
which arises from the ``vacuum'' case in Eq. (\ref{bbsol})  and is called a
DeSitter space.
In simplest terms, a universe undergoing accelerated expansion
grows much bigger in the same amount of time as a universe
undergoing decelerated expansion.  This point is expressed in
Fig. \ref{fig2}

\begin{figure}[ht]
\vspace{1cm}
\epsfysize=5.95cm
{\centerline{\epsfbox{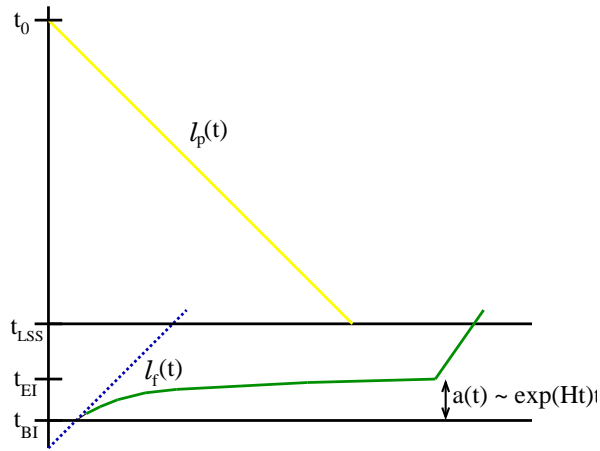}}}
\caption{Lightcone diagram describing the horizon problem
and its solution.
The lightcone $l_p$ (light solid line) shows the maximum region
from the past from which an observer today, $t_0$, can receive
information.
The lightcone $l_f$ shows
the maximum region a light
signal can cover since the creation of the universe
under Big Bang evolution (dashed line) and under inflationary
evolution (heavy solid line).}
\label{fig3}
\end{figure}

This simple idea of having an accelerated growth of the
scale factor solves both cosmological puzzles.  In the case
of the horizon problem, Fig. \ref{fig3} illustrates the
solution.  Before addressing this solution, it is
important to understand the significance of the time denoted $t_{LSS}$ 
in this figure.  This denotes the time after which
photons effectively stop interacting with all forms of matter,
an event called ``last scattering''.
Using basic particle physics it is possible to calculate the
interaction rates of photons with all other particles,
and from that determine quantitively when last scattering occurs.
This is an important era in the history of the universe.
After this time, all the photons in the universe are basically
not interacting with anything.  This means whatever the
distribution of the photon field was at time of
last scattering, it remains intact subsequently, aside from
simple and calculable effects from the expansion of the universe.
Thus, the CMB photon field that is observed today is simply
a snapshot of the field at time of last scattering.
This is the oldest fossil remnant from the early universe that we have
available to us today and that is why there is intense interest
in understanding every aspect of the CMB.

Returning to Fig. \ref{fig3}, the dashed line
$l_f(t)$ represents the forward light-cone
of a photon,
as it would evolve in the Big Bang model,
emitted at the beginning of the universe up to
the time $t_{LSS}$.
The region to the left of $l_f(t_{LSS})$
at $t_{LSS}$ represents the region
that has been in causal contact at some point in
the past before $t_{LSS}$.  This is the maximum  region that
the Big Bang model predicts in which causal processes could
for example produce a uniform temperature distribution.
The line $l_p$ is the past light-cone of an observer today, $t_0$.
The horizon problem in this context is that 
in a Big Bang evolution, the region
of universe at $t_{LSS}$ from which information is received today
is much larger than that which could have been in causal contact
at time $t_{LSS}$.  In fact simple
calculations tell us that there would be something
to the extent of $10^{5}$ causally disconnected regions at time
of last scattering from which we receive information.
This number places into perspective just how starteling it
then is that observation finds the CMB temperature
basically is the same up to parts in $10^5$ from all
directions in the sky.  The inflation solution to
this problem is simply that in such a expansion period, the
forward light cone grows much faster.  For example for
an exponentially growing scale factor, the solid line $l_f$
in Fig. \ref{fig3} illustrates
the exponential growth of the forward light-cone,
so that at time of last scattering, the region of causal
contact incorporates the region within
the past lightcone of an observer today, thus solving
the horizon problem. 

Inflation also solves the flatness problem.  The key point being that
in a universe undergoing accelerated expansion, the flat geometry becomes
an attractive point.  For example, for an exponentially expanding
scale factor, the universe tends exponentially
close to a flat geometry.  Thus inflation provides a dynamical
mechanism for making the universe extremely flat in its early
phases.

\begin{figure}[ht]
\vspace{1cm}
\epsfysize=5.95cm
{\centerline{\epsfbox{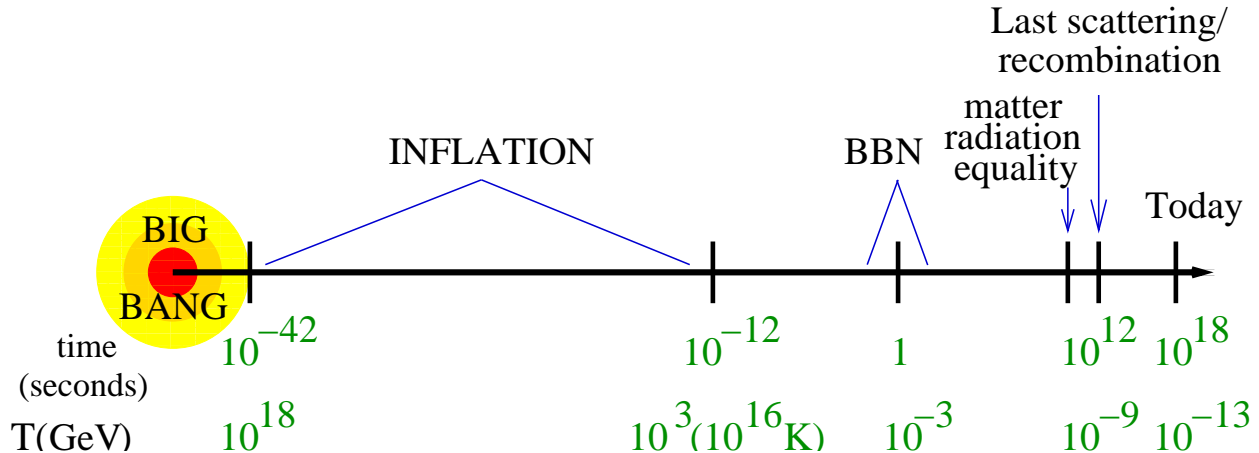}}}
\caption{A time history on the universe}
\label{fig4}
\end{figure}

Inflation is pictured to occur at an early
epoch of the universe, although the time period is not
very well pinpointed.  In Fig. \ref{fig4} a time history
of the universe is given based on the Big Bang model, with
some key events indicated.  The success of the Big Bang model
since nucleosyntheis (BBN) to the present suggests that the necessary
amount of inflation needed to solve the cosmological puzzles
could not have occurred during this recent time period.  
Moreover by combining knowledge from particle physics and cosmology,
the time can be pushed back substantially.
Supersymmetry (SUSY) is an idea that has an important role to
inflation model building.  The details of SUSY and its relevance
to inflation
will be discussed later in the article.  Here
it is only mentioned that we know SUSY must no longer be
a symmetry of nature below
the electroweak scale ($T \sim {\rm TeV}$),
Thus if SUSY is to be useful for inflation model building,
than this implies inflation must have occurred somewhere
above the $1 {\rm TeV}$ scale, which from Fig. \ref{fig4}
means occurred before $\sim 10^{-12}$ seconds after the creation
of the universe.

The idea of cosmological repulsion should be credited
to Einstein in 1917 who introduced what he called the cosmological
constant into his General Relativity equations.
His motivation was to allow a static universe in which
this repulsive term counteracts the gravitational attractive force
arising from the matter in the universe.
This idea was dropped after Hubble's discovery that
the universe is expanding.
Subsequently De Sitter in the mid 1920's built a model
involving just a cosmological constant,
which meant the universe expanded exponentially
forever.
In more modern times, in the 70's Gliner was one of
the first to recognize the inflation idea
\cite{gliner}.  Also in the '70's, the idea of cosmological
phase transitions was developed, with major work done
by Kirzhnits and Linde \cite{kl}.  In '79 Starobinsky came up with a model
of exponential expansion using quantum gravitational
effects and higher derivative curvature terms \cite{star}.
The pivotal step was in 1981 with Alan Guth's seminal paper \cite{guth}
which suggested exploiting a stage of exponential expansion
in some supercooled vacuumlike state, that occurs during
a first order phase transition, to solve the
cosmological puzzles.  It was in this paper that the name inflation
was dubbed.  At about the same time Sato \cite{sato} also had
studied the effects of a first order phase transition
which could provide the necessary conditions for exponential expansion.
Guth's paper offered a very clear and elegant idea, but it had
the problem of making the universe too inhomogenous after inflation.
In 1982 the new inflation scenario was suggested
by Linde \cite{lindeni} and Albreacht and Steinhardt \cite{as}.
This model set the basic paradigm of scalar field driven inflation
and nowadays is regarded as the standard picture of inflation.
In the mid-90s it was recognized that since inflation models 
are interacting dynamical systems, they
should exhibit fluctuation and dissipation effects.
This observation led to the warm inflation picture \cite{wi,wi2}.
Although warm inflation is just a completion of the
original inflation picture, it does alter the basic
picture and so it is convenient to treat
as a second dynamical realization of inflation.

\section{Inflation from quantum field theory}
\label{sect6}

For theorists, the reason so much excitement and interest centers
around inflation is that particle physics contains the basic
ingredients necessary to realize inflation.
Our most fundamental understanding of nature at present is
quantum field theory.  In this approach, each particle
is represented through a quantum field.  The dynamics of the whole
system of particles is  expressed through a Lagrangian,
from which the Lagrange equations of motion for all
the quantum fields is obtained.
Current quantum field theories of matter contain three types of
fields, spin $1/2$ fermions, such as the electron,
spin $1$ gauge bosons, such as photons, and spin zero bosons,
such as the Higgs.  Although there is ample evidence
for the existence of
spin $1/2$ fermions and spin $1$ gauge bosons, no
elementary spin $0$ scalar particle has yet been detected.
Nevertheless, all modern particle models rely on the existence
of such fields and the Standard Model
Higgs particle is the foremost candidate that
particle physicists hope to find at the upcoming 
Large Hadron Collider (LHC).

From General Relativity we know that in order to realize
inflation, it requires an equation of state $p < -\rho/3$,
thus a substance with negative pressure.
As it turns out, scalar fields can provide such an equation of
state.  The energy and pressure density of a scalar field is
\begin{eqnarray}
\rho & = & \frac{{\dot \Phi}^2}{2} + {\rm V}(\Phi) + \frac{(\nabla \Phi)^2}{2a^2}
\nonumber \\
p & = & \frac{{\dot \Phi}^2}{2} - {\rm V}(\Phi) - \frac{(\nabla \Phi)^2}{6a^2},
\label{rhop}
\end{eqnarray}
so that a state which is dominated by the potential energy of
a scalar field has a negative pressure.
Thus the idea that has generally been adopted in realizing
inflation from particle physics is to get the potential
energy of some scalar field to dominate the energy density
of the universe for some short time period in the early
phases of the universe, thereby generating the requisite
amount of inflation necessary to solve the cosmological puzzles.
Then once enough inflation has occurred, somehow put the
universe back into a radiation dominated hot Big Bang regime.
The scalar field that performs the task is driving
inflation is called the inflaton.
There are two essential roles the inflaton must perform.
First it must supply an appropriate energy density to be
conducive for inflation. Second the fluctuations of the inflaton
must have the appropriate features to seed density fluctuations in
the universe.

\section{Two dynamical realizations of inflation}
\label{sect7}

\begin{widetext}
\begin{figure}[ht]
\vspace{1cm}
\epsfysize=8.50cm
{\centerline{\epsfbox{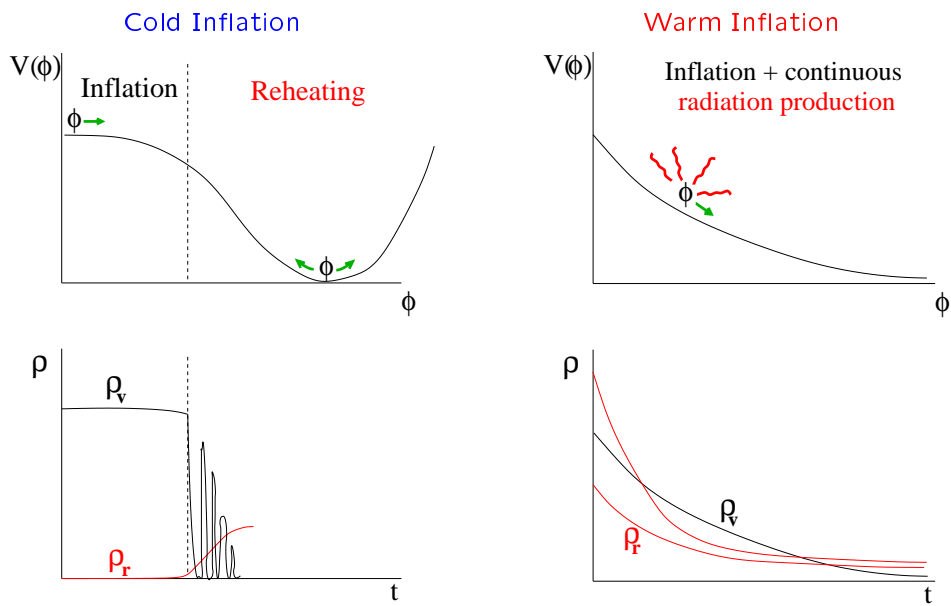}}}
\caption{Comparison of the cold and
warm inflationary pictures, top graphs show the scalar field
evolution and the bottom graphs show the vacuum and radiation
energy density evolution}
\label{fig5}
\end{figure}
\end{widetext}

Inflation is a paradigm, not a theory.  Thus, it does not make specific
predictions in the same way as the Standard Model of particle physics.
Each specific model of inflation makes definite predictions, but the
whole class of models can be tested only by looking for
generic features that are common to all models.
There are two underlying dynamical realizations of inflation,
into which all models fall, cold and warm inflation.
The basic pictures describing both these dynamics is summarized in
Fig. \ref{fig5} and will be discussed in detail in this section.
Cold inflation is synonomous with the
standard inflation picture.  This picture will be
briefly reviewed, since many of its features are also found
in the warm inflation picture.  After reviewing
cold inflation, attention turns to warm inflation.

\begin{figure}[ht]
\vspace{1cm}
\epsfysize=5.95cm
{\centerline{\epsfbox{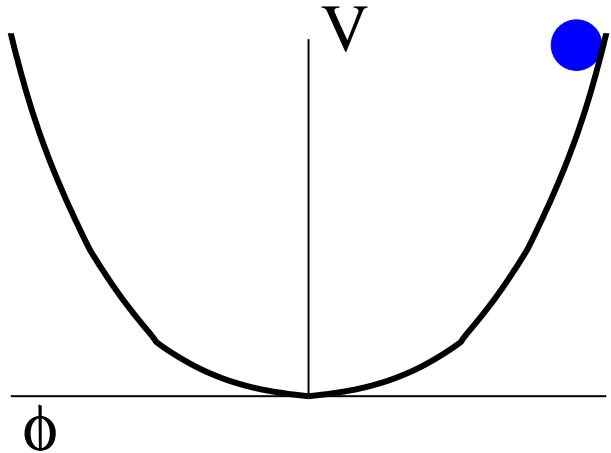}}}
\caption{A typical inflationary potential with the inflaton
initially starting at some large amplitude}
\label{fig6}
\end{figure}

\subsection{cold inflation}
\label{subsect7a}

In the cold inflation picture, the scalar inflaton field is assumed to be
essentially in isolation, thus interacting with nothing else
besides gravity, during the inflation phase.  Also, the scalar field
must carry a large potential energy.
For example, Fig. \ref{fig5} shows a symmetry breaking potential
for cold inflation, in which when the inflaton is just near the top,
$\langle \Phi \rangle \equiv \phi \approx 0$, inflation is expected to occur.
For another example,
Fig \ref{fig6} shows a $\lambda \phi^4$ potential,
in which when the inflaton amplitude is displaced to some
$\phi > 0$, inflation occurs.
The evolution of the scalar field in the
FRW universe is described by the General Relativistic version
of the Klein-Gordon equation
\begin{equation}
{\ddot \phi} + 3H {\dot \phi} 
- \frac{1}{a^2(t)} \nabla^2 \phi 
- \frac{\partial V}{\partial \phi} = 0.
\label{cieom}
\end{equation}
In this equation the Hubble damping term, $3H {\dot \phi}$,
formally acts like a friction term that damps
the inflaton evolution. However this $3H {\dot \phi}$ term
does not lead to dissipative energy production,
since its origin is 
from the coupling of the scalar field
with the background FRW metric.
As mentioned earlier, the inflaton must fulfill both
requirements of driving inflation and providing the seeds
for density fluctuations.  This section only focuses
of the first of these requirements, with the latter taken
up in the next section.

In order for inflation to occur, the inflaton must be potential 
energy dominated, which means the potential energy $V(\phi)$
must be larger than the gradient energy $(\nabla \phi)^2/2$ and the
kinetic energy ${\dot \phi}^2/2$.
Moreover, in order to obtain enough inflation, these
conditions
must persist for some span of time.
This is often referred to as the ``slow roll'' conditions.
To achieve this, the inflaton
field must start out almost homogeneous and almost at rest
in some small patch of space.
This is possible if the 
Hubble damping term, $3H {\dot \phi}$,
is large and the potential is very flat.
In particular, the slow roll conditions require 
$9 H^2 > V''$, which if one considers the analogy to the damped
harmonic oscillator implies an overdamped evolution for
$\phi$, and $V > {\dot \phi}^2/2, (\nabla \phi)^2$/2, so that the
potential energy dominates.
If these conditions hold in a region of space, then
inflation can happen.

For the inflaton dynamics of Eq. (\ref{cieom}), it is instructive
to see how an inflationary scale factor growth occurs.
To a good approximation, the slow roll regime yields an
energy density with the vacuum form in Eq. (\ref{bbsol}).
In particular, the only relevant field during inflation in this
picture is the inflaton and in the slow roll regime 
its equation of state from Eq. (\ref{rhop}) reduces to
$\rho \approx V(\phi) \equiv \rho_V$ and $p = -V(\phi)$, with 
$V(\phi) \approx \ const \ \equiv V_0$.
Applying this to the scale factor equation (\ref{scalee}),
it leads to ${\ddot a} = 8 \pi G V_0 a/3 > 0$, thus
an accelerated expansion or in other words inflation.
This equation is easily solved to yield
$a(t) \propto \exp(\sqrt{8 \pi G V_0/3}t)$, which
is the characteristic exponential growth of inflation described
earlier.
 
The evolution of the energy components in the universe also
can be easily studied in this example.  During inflation
there are two possible components contributing,
the vacuum energy density and the radiation energy density.
For this system the energy conservation equation becomes
${\dot \rho_R} = - 4 \sqrt{8 \pi G V_0/3} \rho_R$, which has
the exponentially decaying solution
$\rho_R \sim \rho_{RI} \exp(-4\sqrt{8 \pi G V_0/3} t)$.
In other words, whatever might be the initial radiation energy
density in the universe,
at the onset of cold inflation this rapidly decays away,
thus supercooling the universe.
As shown in Fig. \ref{fig5}.  during cold inflation,
the vacuum energy density is large and almost
constant, whereas the radiation energy density
is negligible.

Once a region of adequately large potential energy
materializes, the physics
of the subsequent evolution is quite straightforward.
The gravitational repulsion caused by the negative pressure drives
that region into a period of accelerated expansion.
One expresses the amount of inflation
as the factor growth in the scale factor at the end of
inflation $a_f$ to that at the beginning $a_i$, and it is usually 
stated in terms of the efolds $N_e$, where
$a_f/a_i = \exp(N_e)$.  In order to solve the cosmological
puzzles, it is understood that about 60 e-folds of inflation
are needed.

Since in this picture
the inflaton is completely noninteracting with other matter
during inflation, it means in this picture inflation 
dilutes away any particles that are present at the start
of inflation.  Thus whatever the temperature of the universe
is at the start of inflation $T_i$, during inflation it
rapidly falls as $T(t) = T_i a_i/a(t)$.  Since the scale factor
grows by at least the factor $\exp(60)$, effectively this amounts to
a supercooling of the universe, thus the name cold inflation.

Eventually inflation must end and radiation must be
introduced into a very cold universe so as to put it
back into a radiation dominated Hot Big Bang regime.  
In the cold inflation picture, the process that
performs this task is called reheating \cite{reheat}.  In this process,
the Hubble damping term must become small compared to the
${\ddot \phi}$ term in Eq. (\ref{cieom}).  How this arises can
vary from model to model.  In Fig. \ref{fig5} the curvature
of the potential becomes very large at some point,
which then permits oscillations of the inflaton.
In Fig. (\ref{fig6}),
since $H \sim \phi^2$, it means the Hubble damping term will eventually
get smaller than the curvature of the potential $V$ as the scalar
field falls down far enough.  At this point, the inflaton
starts oscillating about the origin.
In both cases the oscillations cause the
inflaton to become kinetic energy
dominated, which terminates inflation.  If at this point the inflaton
is interacting with other matter fields, the oscillations
of the inflaton will lead to particle production so
that, as shown in Fig. \ref{fig5},
the radiation energy density begins to increase.

\subsection{warm inflation}
\label{subsect7b}

The other dynamical realization of inflation is warm inflation.
In this picture, similar to cold inflation, the scalar inflaton field
must be potential energy dominated to realize inflation.
The difference is, in this picture the inflaton is not
assumed to be an isolated, noninteracting field during
the inflation period.  So, rather than the universe
supercooling during inflation, such as in the cold inflation
picture, instead the universe maintains some amount of radiation during
inflation, enough  
to noticeably alter inflaton dynamics.
In particular, the dividing point between warm and 
cold inflation is roughly at
$\rho_R^{1/4} \approx H$, where $\rho_R$ is the radiation energy density
present during inflation and $H$ is the Hubble parameter. Thus
$\rho_R^{1/4} > H$ is the warm inflation regime 
and $\rho_R^{1/4} \stackrel{<}{\sim} H$
is the cold inflation regime. This criterion is independent of
thermalization, but if such were to occur, one sees this criteria
basically amounts to the warm inflation regime corresponding to when $T
> H$. This condition is easy to understand since 
the typical inflaton mass during
inflation is $m \approx H$ and so when $T>H$, thermal fluctuations
of the inflaton field
will become important.

The interaction of the inflaton with other fields implies its effective
evolution equation in general will have terms representing
dissipation of energy out of the inflaton system and into
other particles. A simple phenomenological equation that
expresses this is of the Langevin form
\begin{equation}
{\ddot \phi} + [3H + \Upsilon] {\dot \phi} 
- \frac{1}{a^2(t)} \nabla^2 \phi 
- \frac{\partial V}{\partial \phi} = \zeta.
\label{wieom}
\end{equation}
In this equation, $\Upsilon {\dot \phi}$ is
a dissipative term and $\zeta$ is a fluctuating force. 
Both are effective terms arising
due to the interaction of the inflaton
with other fields.  In general these two terms will
be related through a fluctuation-dissipation theorem,
but details of that would depend on the statistical
state of the system and the microscopic dynamics.
In Sect. \ref{sect8} the origin of these
term from first principles dynamics will be examined.

In order for inflation to occur in this picture, the potential
energy $V(\phi)$ (vacuum energy density)
must be larger than the gradient and kinetic energy,
just as in the case of cold inflation.
In addition the vacuum energy density must be larger than the radiation
energy density $\rho_R$.  This can be seen from the scale factor
equation (\ref{scalee}).  For a universe filled with vacuum and radiation,
this equation becomes ${\ddot a} = 8 \pi G a (\rho_V - \rho_R)/3$,
so that for an accelerating scale factor it requires
$\rho_V > \rho_R$.
From this equation it also can been seen that if $\rho_R$ is just a little
smaller than $\rho_V$, by a factor of two or more, then the
scale factor is fairly unaffected by the radiation component and
the scale factor growth is almost exponential,
just as in the cold inflation case.

The difference from cold inflation is in the evolution
of the energy densities, as can be compared in
Fig. \ref{fig5}.  In warm inflation the radiation
energy does not vanish because vacuum energy
is continuously being dissipated at the rate 
${\dot \rho}_V = - \Upsilon {\dot \phi}^2$.
The energy conservation
equation (\ref{econs}) for this system of vacuum and
radiation becomes 
${\dot \rho}_R =  -4H \rho_R + \Upsilon {\dot \phi}^2$.
In this equation the second term acts like a source
term which is feeding in radiation energy, whereas the
first term is a sink term that is depleting it away. 
Since the rate of depletion from the sink term is proportional to the amount
of radiation present, it means there 
in general will be some nonzero steady state
point for $\rho_R$.  For example, suppose the source term
is just a constant, which is a good approximation
during the slow roll evolution of the inflaton,
$\Upsilon {\dot \phi}^2 = const. \equiv c_0$.
Then the solution to the equation will be 
$\rho_R \approx c_0/(4H) + (\rho_{R0} - c_0/(4H)) \exp(-4Ht)$.
The second term on the RHS of this solution decays away
any initial radiation.  However at large time radiation does not
entirely vanish because of the first term on the RHS.
Thus at large time,
the radiation in the universe becomes independent of
initial conditions and depends only on the rate at
which the source is producing radiation.

As already mentioned, the scale factor equation 
shows that the presence of radiation during inflation
is perfectly well allowed by the equations of General Relativity, since
the only requirement it imposes to realize an inflationary scale
factor growth is that the vacuum energy density is the dominant
component of energy in the universe.   Thus there could still
be say a 10\% or 1\% etc... admixture of radiation, in addition
to the vacuum energy, and inflation would still happen.
This is an important point.  To appreciate it, note that there are
at least five 
scales in inflation,  
the vacuum energy $E_V \equiv \rho_V^{1/4}$,
the radiation energy $E_R \equiv \rho_R^{1/4}$, 
the Hubble scale $H$,
the inflaton mass $m \equiv V''(\phi)$, and
the dissipative coefficient $\Upsilon$. 
In the cold inflation picture, these five
energy scales are related as (i). $E_V > E_R$, 
(ii). $H > m$,
(iii). $m > E_R$, and
(iv). $H \gg \Upsilon$.
Condition (i) is simply a minimal requirement from
General Relativity in order to have inflation.
Condition (ii) is necessary to be in the slow roll regime.
Condition (iii) implies the universe is in a low-temperature
regime where radiation has an insignificant effect on inflaton fluctuations.
Finally condition (iv) implies
dissipative effects have an insignificant effect on
inflaton evolution.

Turning to warm inflation, there are two types of regimes that
must be addressed, weak and strong dissipative warm inflation.
In both these regimes the following  energy scales are the same 
(i). $E_V > E_R$, 
(ii). ${\rm max} \ (\Upsilon,H) > m$, and
(iii). $E_R > m$.
Condition (i) is again required by General Relativity to
realize inflation.  Condition (ii) is the warm inflation
equivalent to being in the slow roll regime.
Condition (iii) implies the inflaton fluctuations are
no longer in a zero temperature state, thus radiation 
will have nontrivial effects on inflaton dynamics and fluctuations.
Finally the last condition, and the one that leads to two
regimes of warm inflation is
(iv). $\Upsilon > 3H$, strong dissipative warm inflation
and (iv). $\Upsilon \leq  3H$, weak dissipative warm inflation.
The notation here is almost self-explanatory, in the
strong dissipative regime, the dissipative coefficient
$\Upsilon$ controls the damped evolution of the inflaton field and
in the weak dissipative regime, the Hubble damping is still
the dominant term.

Even though the presence of radiation need not hinder
inflationary growth, it can still influence inflaton dynamics.
To appreciate this point, an example will
be beneficial.  Consider an inflation occurring at the
Grand Unified Theory scale, which means 
$V^{1/4} \equiv E_V \sim 10^{15} {\rm GeV}$.
In this case the Hubble parameter turns out to be
$H \sim V^{1/2}/m_P \sim 10^{10} {\rm GeV}$.
For cold inflation and weak dissipative warm inflation,
since the Hubble damping term $3H {\dot \phi}$ must 
be adequate to produce a slow roll evolution for the inflaton, 
it also means the inflaton mass 
$m \sim 10^{9-10} {\rm GeV} \stackrel{<}{\sim} 3H$.
The key point to appreciate here is that there are five orders
of magnitude difference here between the vacuum energy scale and
the scale of the inflaton mass.  Thus there is a huge difference in
scales between the energy scale $E_V$ driving inflation and the
energy scale $m$ governing inflaton dynamics.
This means, for example, in order to excite the inflaton fluctuations
above their ground state only requires a minuscule amount
of vacuum energy dissipated at a level as low as $0.001\%$.
This gives a good indication that dissipative effects during
inflation have the possibility to play a noticeable role
(several models of warm inflation exploiting these properties
exist \cite{abinterp,dr,gmn,lf,ml,mb,dj,cjzp,by}).
Of course this is only an energetic assessment that is suggestive
of interesting physics.  It remains a question that a
proper dynamical calculation must answer.  In particular,
the universe is expanding rapidly during inflation at a rate
characterized by the Hubble parameter $H$.  One must
determine whether the fundamental dynamics responsible for
dissipation occurs at a rate faster than Hubble expansion.

The other difference of warm inflation to cold inflation is how dissipation
effects the parameters of the underlying first principles model.
This in particular becomes evident in the strong dissipative regime
when $\Upsilon \gg 3H$.  To appreciate this point,
note that in cold inflation the inflaton motion is damped by
only the $3H {\dot \phi}$ term.  Thus in order to have slow roll
evolution, it requires that the inflaton mass $\sim \sqrt{ V''}$, must
be less than $3H$.  However in typical quantum field theory
models of inflation, it is very difficult to maintain such a
tiny inflaton mass, a point which is explained in greater
detail in Sect. \ref{sect11}.  This problem is often called 
the ``$\eta$-problem'' \cite{Arkani-Hamed:2003mz}.  
In contrast, in warm inflation
slow roll motion only requires $V'' < (3H + \Upsilon)^2$,
so when $\Upsilon > 3H$ it means the inflaton mass can be bigger
than in the cold inflation case, even much bigger.
This relaxation on the constraint in the inflaton mass permits
much greater freedom in building realistic inflaton models,
since this ``$\eta$-problem'' is comfortably eliminated.

Another model building consequence differing warm inflation to
cold inflation relates to the region of the scalar field 
amplitude $\phi \equiv \langle \Phi \rangle$
in which the inflation occurs.
For cold inflation, for the simplest
kinds of potentials, which also are the most commonly used,
such as $V = \lambda \phi^4/4$ and $V = m^2 \phi^2/2$, 
calculations show that the initial
inflaton amplitude has to be above the Planck scale
$\phi_i > m_p$.  This arises because the Hubble damping term,
$3H {\dot \phi}$, in these models increases with larger field amplitude,
so in order to reach a regime in which slow roll 
occurs for an adequate time to yield the desired
60 or so efolds of inflation, it forces
these large field amplitudes.    However from the perspective
of particle physics, and the ultimate goal
of building a realistic inflation model, this condition poses
a problem which will be discussed in further detail
in Sect. \ref{sect11}.  The upshot is that it generally forces 
more complications into the model building, simply to avoid this problem.
On the other hand in warm inflation, when $\Upsilon > 3H$,
the added dissipation means the time period
of slow roll necessary to obtain the desired
60 or so efolds can be achieved
with the inflaton traversing over a much smaller
region of its amplitude, 
thus allowing its amplitude to be smaller.  Although
the exact answer requires calculation, what is found is that
for these simple monomial potentials, 
for warm inflation the inflaton amplitude is
below the Planck scale $\phi < m_p$, which simplifies model
building as will be further discussed in Sect. \ref{sect11}.

\section{Density perturbations}
\label{sect8}

Zeldovich once likened the universe 
to an elephant, indeed \cite{pt85}.
His analogy was that globally the shape of the universe may
be something very odd, even that of an elephant.
Our observable universe in this analogy would be some small
part of the elephant's surface, which would look
reasonably flat, because its just a small local region.
Moreover, in this local region of the elephant's surface,
his skin would have wrinkles, and these would be analogous
to the perturbations in our universe that emerge as
planets, stars, galaxies etc...

As already discussed
inflation can explain the origin of homogeniety, but what
about the small inhomogenieties in energy density 
that the universe has.
Moreover, as pointed out in Sect. \ref{sect3},
observation shows that structure in our universe is correlated on
scales far too big to be explained by the Standard Cosmology.
Inflation has the necessary kinematic features to explain 
the origin of such structure.  To appreciate this point, consider
the origin and growth of fluctuations in an expanding universe.
At a given epoch
of the universe at say cosmic time $t$, causality would imply
correlated structure can be produced at that time up to
a maximum length scale $\sim cH^{-1}(t)$.
Alternatively, if a fluctuation was created at some earlier epoch
$t_e < t$, and if this fluctuation remains unaltered expect growing due to
expansion, then at time $t$ the largest correlated structure
to which this could evolve would have a
length $\sim a(t) c H^{-1}(t_e)/a(t_e)$.
In a Standard Cosmology, meaning $a(t) \sim t^{n}$ with $n < 1$,
$cH^{-1} \sim ct$, so that the largest scale of any correlated structure
whether produced in that epoch or from an earlier
epoch will be $\sim ct$.  Consider how these results
of Standard Cosmology contrast with those for an inflationary
cosmology.  In this case the largest possible scale that can be causally
produced in a given epoch is $\sim c H^{-1} \approx const$,
whereas if the structure was produced at an earlier 
epcoh of inflation $t_e < t$, 
then at time $t$ it is correlated on a scale
$\approx \exp[H(t-t_e)] cH^{-1} > cH^{-1}$.
In other words, the scale of correlations grows exponentially.
Consider the example where a fluctuation is created at the onset
of inflation, then after $N_e$ efolds of inflation
it would have grown by a factor $\exp(N_e)$.
Thus if $N_e \approx 60$ it means the length scale of the fluctuation
has grown by a factor of $\exp(60)$.  Contrast this with how
much the fluctuation would have grown in the same period of
cosmic time in a Hot Big Bang evolution.  The 60-efolds
of inflation took a time $\sim 60/H$. In this time interval
the growth of the fluctuation length scale would have only
been by a factor of $\approx 60$, so much much less.
It is this kinematic feature of inflation which allows
production of structure that is correlated on scales much
larger than conceivable by causality in a
Standard Cosmology.

The kinematic example above implies that 
provided some dynamical mechanism
exists to actually create fluctuations during inflation,
the exponential expansion will provide the means
to expand those fluctuations to very large scales. 
Dynamically, the inflaton provides the necessary means for producing
density fluctuations during inflation.
The key point here is that whereas the homogeneous mode
of the inflaton field drives the background cosmology,
the small wavelength modes of the field can create the
initial density fluctuations.
For a quantum field at either zero or finite temperature,
it is well known fluctuations of the field will be produced.
These fluctuations can be regarded as waves of all
possible wavelengths moving in all possible directions.
For a field governed by a damped evolution as in
Eqs. (\ref{cieom}) or (\ref{wieom}), 
the small wavelength modes will oscillate over time,
so that the amplitude averaged over a substantial period of time
vanishes.  However in an exponential expanding universe
the situation is different.  The wavelengths of all
fluctuations of the field grow exponentially along with
the expanding universe.  When the wavelength of any
particular mode becomes larger than the damping scale,
$cH^{-1}/3$ or $c(\Upsilon + 3H)^{-1}$ in 
Eqs. (\ref{cieom}) and (\ref{wieom})
respectively, that particular mode stops oscillating and
so its amplitude freezes at some particular value.
After this time, termed "freeze-out", the amplitude of this mode
remains almost unchanged whereas its wavelength keeps
growing exponentially.  The outcome is the appearance
of a classical field that does not vanish after averaging over 
macroscopic intervals of space and time.
This process thereby creates the seeds for structure formation.
Over time these initial fluctuations will grow due to
gravitational instabilities.  The regions of mass overdensity
will start attracting more matter to them and those of underdensity
will start collapsing, ultimately leading to the clumped formations
found in the universe today.  Moreover, inflation provides
the ingredients for producing the almost scale free spectrum
of density fluctuations found in the universe.  This arises
since slow roll during inflation implies very little changes in
the properties of the inflaton field during inflation.  Thus
each  mode that freezes out during inflation will have approximately
the same amplitude, thus seeding a near scale free spectrum
of density fluctuations.
 
The discussion so far is common to both cold and warm inflation.
The difference arises in the nature of the fluctuations in these
two dynamics.  In cold inflation, the inflaton produces
zero-point ground state fluctuations.
The amplitude of the fluctuations at the point of freeze-out
are given by \cite{cidp}
\begin{equation}
\delta \phi \sim H.
\end{equation}
In warm inflation, the inflaton is in some excited statistical
state.  The fluctuations are induced on the
inflaton field by the noise force
$\zeta$ in Eq. (\ref{wieom}). 
The most common example, and the only one that will
be detailed here is the case where it is in thermal
equilibrium at a temperature $T > m$.
In this case the amplitude of the fluctuations
at freeze-out are 
\begin{equation}
\delta \phi \sim \sqrt{HT}
\end{equation}
in the weak dissipative regime \cite{bf2} and   
\begin{equation}
\delta \phi \sim \sqrt{T(H \Upsilon)^{1/2}}
\end{equation}
in the strong dissipative regime \cite{abnp}.

The key point is that the nature of the fluctuations in cold
versus warm inflation are qualitatively different.
This opens up the possibility that observational signatures 
could be found that might distinguish these two dynamics.
This will be discussed further in Sect. \ref{sect12}.

\section{First principles dynamics}
\label{sect9}

It has already been pointed out in Sect \ref{sect7} that even a little
conversion of vacuum energy into radiation can have significant
effect during inflation.  Also the phenomenological equation (\ref{wieom})
has been presented as an effective evolution equation for the
inflaton field once interactions with other fields
are integrated out.  The question that remains to be
answered from quantum field theory is whether both these
effects  actually occur and if so in what sorts of models.
To address these questions, our task in this section
is to start with the fundamental
Lagrangian and from that derive the effective equation
of motion for the scalar inflaton field.  

For any inflaton model
the basic Lagrangian quite generally has the form 
$L = L_S + L_R + L_I$. Here $L_S$ is the Lagrangian
of the inflaton system itself, which has the general form
\begin{eqnarray}
{\cal L}_{S} =
\frac{1}{2} {\dot \Phi}^2 -
(\nabla \Phi)^2 
- V(\Phi) .
\label{Lphipsi}
\end{eqnarray}
In any inflation model, this inflaton must interact with other fields,
since there must exist channels from which 
the vacuum energy contained in the inflaton
field ultimately can be released into radiation energy in order to
end inflation and put the universe back into a Hot Big Bang
evolution.  These interactions are contained in the $L_I$ part of
the above Lagrangian.
The question is whether this conversion process
occurs exclusively at the end of inflation, as pictured
in cold inflation, or does it occur concurrent with
inflation, as pictured in warm inflation.
The most common sorts of interactions would be the inflaton
coupled to bosonic fields such as $0.5 g^2 \Phi^2 \chi^2$
or fermion fields as $h \Phi {\bar \psi}{\psi}$,
i.e. $- L_I = 0.5 g^2 \Phi^2 \chi^2 + h \Phi {\bar \psi} \psi$.
Finally $L_R$ contains all other terms associated with all fields aside
from the inflaton, like the
$\chi$ and $\psi$ fields in this example.

For this Lagrangian, the quantum operator equations of
motion can be immediately written down, one for
each field.   These equations in general are coupled to each
other through nonlinear interactions.  We are interested in the
evolution equation of the fields, or actually their expectation
values, given the initial state of the system at some initial
time $t_i$. 
For the inflation problem, more specifically
we are interested in obtaining the effective equation of motion
for the scalar inflaton field configuration 
$\phi \equiv \langle \Phi \rangle$, after
integrating out the $\Phi$ quantum fluctuations, and the effects of
all other fields with which $\phi$ interacts, like the
$\chi$ and $\psi$ fields in $L_I$.
This is a typical ''system-reservoir''
decomposition of the problem, as familiar in
statistical mechanics \cite{weiss}.  In our
case $\phi$ is the
system and all other dynamical degrees of freedom
constitute the reservoir.

\begin{figure}[ht]
\vspace{1cm}
\epsfysize=5.95cm
{\centerline{\epsfbox{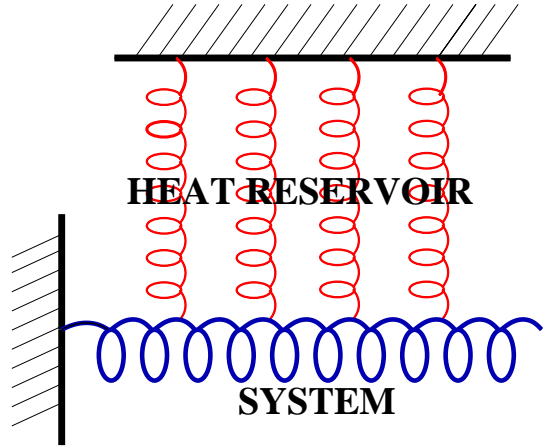}}}
\caption{A mechanical example of a system-reservoir configuration}
\label{sysres}
\end{figure}

A simple mechanical model illustrating the system-reservoir
procedure is shown in Fig. \ref{sysres}.
In this figure the system is the  big horizontal
spring.  This big spring is coupled to a lot of
small springs, which comprise the reservoir.
The equations of motion for all these springs can be written
down.  The solutions of the equations for the small reservoir
springs can be solved as a function of the big spring coordinate.
These solutions can be plugged into the big spring's equation of motion.
This procedure in the system-reservoir language is described
as integrating out all reservoir degrees of freedom
(small springs),
to arrive at an effective evolution equation for the
system (big spring).

The system-reservoir approach is applied to many problems in physics.
It is instructive to state a few examples here. 
A very common example is Brownian motion,
where one is interested in the evolution of one
particle that is singled out, which is immersed in a fluid
and interacts with particles in that fluid.  One seeks an 
evolution equation for this Brownian particle,
once the effect of all the other particles are integrated
out and represented through effective terms in this
evolution equation.
The analog of the Brownian particle in our problem is the background
field $\phi$ and
the reservoir bath in our example contains the $\phi$ fluctuation modes, 
the scalar $\chi$ and the spinor
$\psi$.
In condensed matter
physics, the system-reservoir approach is used in a range of
situations, such as for example 
in the tunneling of a trapped flux in a SQUID,
interaction of electrons with polarons in a metal and
in Josephson junction arrays \cite{weiss,cmsr}.

The procedure for obtaining the $\phi$ effective
equation of motion is first to replace 
the field $\Phi$ in the Lagrangian by
$\Phi = \phi + \sigma$, where $\langle \Phi \rangle \equiv \phi$
and $\sigma$ are the quantum fluctuations
of the $\Phi$ field.  The equation of motion for
$\phi$, with potential for example $V= m^2 \Phi^2/2$, then becomes
\begin{eqnarray}
\ddot{\phi} + 3H{\dot \phi}(t) 
+m^2 \phi(t) - \frac{1}{a^2(t)} \nabla^2 \phi
+ g^2 \phi \langle \chi^2 \rangle + g^2 \langle \sigma \chi^2 \rangle
+ h \langle {\bar \psi} \psi \rangle = 0
\label{vphiave}
\end{eqnarray}
In principle what one is trying to do now is solve the
quantum operator equations of motion of all the other fields, i.e.
$\sigma$, $\chi$ and $\psi$, as a function of $\phi$,
substitute these above in Eq. (\ref{vphiave}), 
take the specified expectation values
and what would emerge is the sought after effective evolution
equation for $\phi$.  In practice this procedure can not
be done exactly, but various perturbative and resummation methods 
are used.  It is not the purpose of this
review to explore these approximation methods,
however the interested reader can examine the
following references \cite{bgr,fpdis,lawrie}. 
Here only a few general features
of the effective $\phi$ equation of motion are mentioned.
First since $\phi$ is isolated, it becomes an open system,
so one should expect the $\phi$ effective 
equation of motion to be nonconservative.
Second, the fields $\chi$, $\psi$ etc...  at a given time
$t_0$ in general will be functions of $\phi$ at
all earlier times $t < t_0$.  Thus the expectation
values $\langle \chi^2 \rangle$ etc... in
Eq. (\ref{vphiave}) will be nonlocal in time
with respect to $\phi$, which is consistent with
the first general fact, since this will lead to a
nonconservative equation.

For the reader familiar with the effective potential in
Lagrangian quantum field theory,
the origin of the $\phi$ effective equation of motion also can be
understood heuristically in another way. 
An effective potential calculation  
applies to the situation where $\phi$ 
is a static background.
Its interaction with the other quantum fields leads to
the creation of
quantum fluctuations which are emitted off $\phi$, propagate
in space and time, and then are reabsorbed by $\phi$.
These processes are known in the subject as loop corrections, which modify the classical
potential and lead to the effective potential.
Now suppose $\phi$ is not in a static situation,
that it is changing in time.  In this case the same loop
corrections mentioned above would occur. However at the time
of emission and absorption, the state of $\phi$
has changed.  Thus these loops no longer simply modify
the potential of $\phi$, but also introduce terms
which mix products of $\phi$ at different times,
thus introducing temporally nonlocal terms into
the $\phi$ evolution equation.

The key issue then is, given a particle interaction structure
in the Lagrangian, what types of
dissipative effects this leads to during inflation.
One particular interaction structure has shown to yield
very robust radiation production during inflation. 
This is a two stage mechanism,
involving the inflaton field coupled to a heavy scalar field
$\chi$ which in turn is coupled to light fermion
fields $\psi$ as 
$g^2 \Phi^2 \chi^2 + h \chi  {\bar \psi}^{\chi} \psi^{\chi}$ \cite{brpl1}.
In this case the background inflaton field $\phi$ acts
as a time dependent mass to the $\chi$ field. As $\phi$
changes over time, the $\chi$ mass changes, thus altering the
$\chi$ vacuum.  This leads to virtual $\chi$ production which
then decay into real $\psi$ particles.
This type of interaction structure is very common in 
particle physics models.  Thus many models of inflation,
which previously were thought to be exclusively cold
inflationary are now known also to have regimes
of warm inflation \cite{brpl1,hm,bb}.

\section{Particle physics  and cosmology}
\label{sect10}

\begin{figure}[ht]
\vspace{1cm}
\epsfysize=7.00cm
{\centerline{\epsfbox{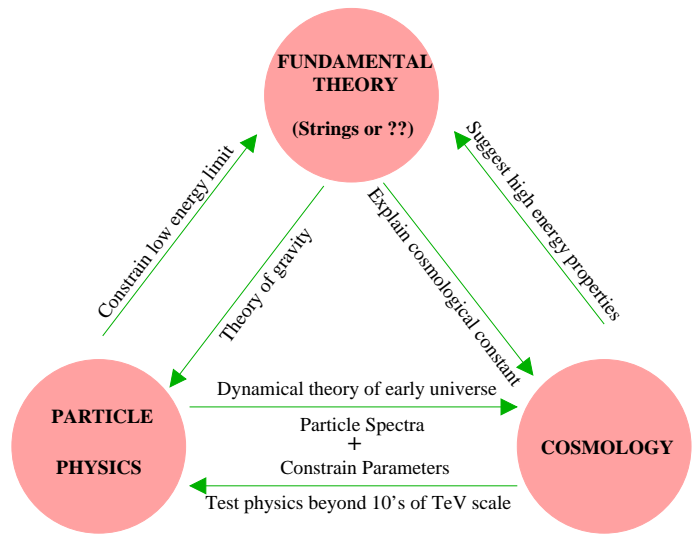}}}
\caption{The three main areas of fundamental physics today.
The arrows connecting the three areas show how each
field assists and is assisted by the other two.} 
\label{fig7}
\end{figure}

One might feel that inflation is a nice idea, since
it elegantly solves the cosmological puzzles,
but because it occurs so far back in the past, what use could
it serve physics today.  To answer this concern,
first some understanding must be developed between
the interplay of cosmology and particle physics.
At first glance it may appear awkward that there should be any
significant connection between cosmology, which
treats the largest scales in the universe, and particle
physics, which treats the smallest scales.  However due to
the expansion of the universe, it means in the past
the universe was hotter and denser, so that short scale
physics becomes increasingly important.
This fact has been appreciated since the early days of
the Big Bang model.  In recent times,
the huge increase in precision measurements in cosmology
combined with the realization that every GeV of
center-of-mass energy gained at
colliders means a long, hard and costly effort, have given impetus
to the union of these two disciplines.
There are inherent ideological differences between these two subjects.
Cosmology is an observational science, not an experimental one.
There is just one data sample, our universe.
The initial conditions for it have been preset and not only
are they not
at our disposal to adjust but we also do not know what
exactly they were.  All information that we can learn
about cosmology comes only from what we can observe from
the presentday universe.
Theory plays the role of providing
the basic paradigms within which observations are framed.
In contrast particle physics is an experimental science.
One can set up a large number of different controlled
tests in the laboratory and observe first hand their
outcomes.
The most basic question that researchers involved in
the interface between the two fields are trying to
answer is how best to take advantage of the differences
to further understanding in both disciplines.

Fig. \ref{fig7} gives an overview of the state of theoretical
particle physics today. 
The bottom two bubbles represent these two disciplines
and show how today they are both active areas of exchange between
theory and experiment.  The top bubble represents the idealized
final theory, that at present is said to be in the process
of getting synthesized by the interplay of these two disciplines.
The arrows connecting the bubbles represent how each field
assists and is assisted by the other two fields.
Thus in the case of cosmology, its role to particle physics
is to provide a testbed for physics at very high energy scales
above the 10s of TeV level, scales too high to be studied
in the conceivable future in a man made
collider.  In return particle physics is
providing the means for determining a dynamical theory
of the early universe which could be tested
by cosmology.

Falling short of the final theory of everything, one of the
more intermediate goals in particle physics is to determine
the dynamical theory of the early universe.
In particular, the goal is
to build a quantum field theory
model which reduces to the Standard Model at low energy
$E < 1 {\rm TeV}$ and explains the observed phenomena
of particle cosmology at high energy scales.
Inflation is the centerpiece of this endeavor.
It has already been an excellent example of the interplay
between these two disciplines.
Realizing inflation from a particle physics
models has its challenges.
It is not as simple an issue as putting in
a scalar field into a particle physics model which has the appropriate
features to drive inflation and produce observationally consistent
density perturbations.  One can build many toy models
of inflation that do these two tasks.
In coupling with particle physics, additional
questions arise such as how does this scalar field
interact with other fields in the model and is there
any motivation, such as through symmetry, for this field.
These questions in turn lead to consequences for the
final temperature of the universe at the end of inflation,
and how that affects the subsequent Big Bang physics,
such as through production of relics, magnetic fields, baryon asymmetry,
and defects.
Also, the scalar field may lead to other consequences for
particle physics at collider scales.  Finally,
a serious treatment of particle cosmology also
must quantify other phenomena such as the
baryon asymmetry, the dark matter content, magnetic
fields in the universe, and cosmological constant to name
an important few.
In building such a model, particle physics tends to
adopt a minimalist attitude, by trying to restrict the number of
free parameters in the model and find relations between
different phenomena.  One of the ideas of constructing a dynamical
model is the hope that different phenomena are found to
be interrelated in some way.  Thus once one adds all these
requirements, the goal of building a QFT model
incorporating particle cosmology and particle physics becomes
a respectable challenge.

\section{Model building aspects}
\label{sect11}

The previous section set out the big picture and there are
many directions in which that could be developed.
Here we return to our discussion of warm inflation and address
the special model building features that this dynamics 
has which distinguish it from cold inflation.  Before  going into
that, first some background into the basic ideas of model building
in particle physics is reviewed.  
The idea of building a particle physics model boils
down to choosing a particular collection of
spin $0$, $1/2$ and $1$ fields, and from these constructing
a Lagrangian with some particular choice of interaction
terms amongst all the fields. Two guiding concepts
play important roles in this construction, symmetry and
renormalizibility.

To any transformation acting on the fields
that leaves the Lagrangian invariant,
there corresponds a symmetry.
Such symmetries can be global, meaning they simply
shuffle fields amongst themselves, or local,
meaning the transformation is different at every
point in spacetime. There are also symmetries involving
transformations of spacetime.  All these symmetries
have important impact on an inflation model, in determining
the interaction structure of the inflaton field, which
in turn determines the effects discussed in previous sections,
such as dissipation, fluctuation, radiation content etc...  
Another important type of symmetry is supersymmetry,
which is the idea that
the basic laws of nature at the fundamental level
are invariant if bosons and fermions are interchanged in
the Lagrangian (for a simple review see \cite{gk}).  
Such a symmetry, quite amazingly is consistent
with relativistic quantum field theory.
In particle physics supersymmetry has been important for
solving the hierarchy problem.  This problem amounts to
the fact that
when there are divergent quantum loop corrections,
the scale of any dimensional quantity in the theory
is not stable.  For example, loops introduce divergent quantum
corrections to masses.  To maintain the scale of these masses
then requires fine tuning of parameters at each order in
perturbation theory, which is clearly not a meaningful procedure.
One of the key features of SUSY is that when one calculates
these quantum loops contributing to the masses or any other quantity,
the particles and their superpartners
both enter the loops.  A well known fact of quantum field theory
is that the sign of fermion loops is opposite to that of
boson loops.  Thus for superpartners with the same masses
and couplings, which would be the case due to
the supersymmetry, these loop corrections would
cancel.   For this reason, in a supersymmetric theory,
the masses and any other quantity 
can be stable to quantum corrections,
This feature of supersymmetry is also important for
inflation model building.  Density perturbation
and slow roll constraints
require that the inflaton potential be very flat.
If the inflaton is interacting with other fields,
then as already mentioned, there will be quantum loop
corrections to the inflaton effective potential.
However in a SUSY theory these loop corrections cancel 
to an adequate degree to render the 
tree-level flatness of the potential
to be preserved. 
For this reason, in a supersymmetric theory,
the inflaton potential can be stable to quantum corrections,

The second important element of model building mentioned
above was renormalizability.
A typical quantum field theory calculation involves the
summing up of quantum fluctuations arising from the various
fields, i.e. the quantum loops mentioned above. 
These fluctuations emerge at all energy scales
up to some maximum upper bound.  If the upper bound
is infinity, then it is possible that these summations
over quantum fluctuations can be divergent.  This is a typical
occurrence in quantum field theory calculations, and to treat
these infinities, the procedure of renormalization is needed.
In this procedure, additional terms are introduced into
the Lagrangian, which themselves become infinite
as the upper cutoff is removed, but which
exactly cancel the infinities from the loop integrals.
Depending on the type of interactions in the Lagrangian,
the number of such counterterms needed to cancel all
loop divergences could be finite or infinite.

In the older days, quantum field theory was regarded as a fundamental
description of nature at all energy scales.  This meant
the theory had no upper energy cut-off and so such loop
infinities generally would appear.  Thus it was argued that the
only sensible quantum field theories were those that required only
a finite number of counterterms to cancel all loop divergences;
such theories have historically been referred to as renormalizable.
This condition imposed severe restrictions on the types of
interactions in the Lagrangian.  For example, fermions could
only interact via Yukawa couplings as $\Phi {\bar \psi} \psi$
and scalar fields interaction terms could involve at most
a product of four fields.
Nowadays Lagrangian quantum field theories are viewed not as
fundamental but rather effective theories that are simply low energy 
approximations to some more fundamental theory such as strings.
This modern viewpoint frees the
restrictions on the interactions to not just the renormalizable
ones, since the Lagrangian theory in any event is only
meant to be valid up to some upper energy cutoff.

Collider experiments have convincingly shown 
that Lagrangian quantum field theories,
in particular the Standard Model, is valid at least up
to the 1 TeV energy scale.  Thus the upper energy cutoff is generally
believed to be somewhere above this scale, although where exactly
is a matter of differing opinions.  The most typical scale is thought
to be the Planck scale $m_P \approx 10^{19} {\rm GeV}$, above which
quantum gravity effects are believed to become significant.
In the effective theory approach, the relevance of a cutoff means the
quantum field theory will have nonrenormalizable interaction terms
that can become important above the cutoff scale.   
Typically very little other information is available, so
one must simply write all possible such terms, usually
an infinite number of them.  For example for a scalar field $\Phi$,
these nonrenormalizable terms quite generally would be
expressed in a series like $\sum_{n=1}^{\infty} a_n \Phi^4 (\Phi/m_P)^n$.

For inflation, these nonrenormalizable interactions would be
important when $\langle \Phi \rangle \equiv \phi > m_P$.  
In fact for model building
this large $\phi$ amplitude regime is a disaster, due to to the
infinite number of unsuppressed terms in the Lagrangian.
For this reason, inflation models must be restricted to
the regime $\phi < m_P$.  This is an important point,
because, in the simplest sorts of inflationary
potentials, meaning monomial potentials
like $m^2 \Phi^2$ or $\lambda \Phi^4$, 
as mentioned earlier, in order to achieve adequately
large Hubble damping from the $3H {\dot \phi}$ term,
it forces a very large field amplitude
which calculations of cold inflation show are in the region of
$\phi > m_P$ \cite{ci,Arkani-Hamed:2003mz}.  
As such, from a model building
perspective, these are not valid models.
In contrast, for warm inflationary dynamics in the strong
dissipative regime, the increased dissipation
allows smaller field amplitudes which 
calculations show for the simple monomial
potentials lead to inflation with $\phi < m_P$,
thus model building is consistent \cite{abnp,bb}.

Nonrenormalizable interaction terms introduce another possible
hurdle for model building.   In cold inflation, since the damping
term is $3H {\dot \phi}$ with $H^2 \sim V/m_P^2$, in order for
slow roll to occur, it means the curvature of the potential must be
less than the Hubble scale $V'' < 9H^2$.  However
supersymmetry models also typically can have 
nonrenormalizable interaction
terms of the form
$V\Phi^2/m_P^2 \sim H^2 \Phi^2$ \cite{Arkani-Hamed:2003mz}.
In cold inflation such terms would spoil the slow roll conditions,
and so additional constraints on model building are introduced.
In contrast, in strong dissipative warm inflation,
since the damping term is now $\Upsilon > 3H$, slow roll is possible
with much bigger curvature of the 
potential $\Upsilon^2 > V'' > 9H^2$ \cite{abnp,bb}.
Thus these sorts of nonrenormalizable interactions have little
consequence.

These differences arising from the warm and cold dynamics,
lead to significant differences in the model building prospects
in the two cases.  In general for particle physics models
such as the Standard Model and its extensions, at high energy scales
it is easy to identify  monomial scalar potentials, thus potentials
conducive for warm inflation.  However, for cold inflation, to overcome
the large $\phi$-amplitude problem, the simplest models,
known as hybrid models, require two scalar fields coupled
in a specific way.  Such models do not readily arise in particle
physics models but must be added on,
thus introducing more parameters and in turn decreasing
the predictability of the model. 

These model building issues are particularly timely.
In two years time the next generation of colliders is expected to
be commissioned, the Large Hadron Collider (LHC), which is under
construction
at the Swiss-French border near Geneva at the European
laboratory for particle physics CERN.
This machine will accelerate protons to very high energies
and then bring them into collision at a center of mass energy
of 14 TeV.  Over a time period spanning beyond the past 10 years,
evidence from collider experiments at
lower energies to the LHC combined with theory have led to
some highly awaited discovery prospects at the LHC.
This list starts foremost with the expectation to detect the
Higgs boson.  Beyond that, or in some quarters maybe
instead of that, the most common phrase being touted is  
finding physics beyond
the Standard Model.  In particular the top
of this list of anticipated
outcomes is the discovery of supersymmetry.

SUSY must be a broken symmetry in our world, since we do not observe
any of the superpartners.  For example, in supersymmetry there
is a spin-zero particle with the mass and electric charge of  
the electron.  If such a particle existed many experiments
would have already detected it.  What we expect is that
at sufficiently high energy SUSY becomes unbroken.
There is no firm knowledge of what this energy scale is,
since in fact as of today there is no firm knowledge that
SUSY really has anything to do with nature at all.
There are convincing theoretical arguments in
favor of its existence.  Moreover, Higgs physics
and its relation to the masses of the Z and W bosons,
offer arguments that SUSY may be restored at around the
TeV scale.  Experiments at Fermilab of protons colliding
at center-of-mass
energy $\sqrt{s} = 2 {\rm TeV} $ and at the 
Large Electron Positron Collider
LEP at $\sqrt{s}= 200 {\rm GeV}$
so far have shown no evidence for the existence of SUSY.
The LHC will have almost an order of magnitude increase
in center-of-mass energy, which offers a significant new
testing ground for SUSY.  

Particle physicists 
have constructed extensions to the Standard Model
that incorporate SUSY.  The simplest such extensions are the
Minimal-Supersymmetry-Model (MSSM) and
the Next-to-Minimal-Supersymmetry-Model (NMSSM),
with the names being self-explanatory.  The interesting
point for warm inflation is it has been shown that in the
NMSSM, the same scalar field responsible for the Higgs mass
also can be the inflaton \cite{bb}.  At large amplitudes
the potential for this scalar field has the
form $\lambda \Phi^4$. Thus cold inflation would
not be consistent, due to the large amplitude problem,
but warm inflation is possible.
Realizing inflation is one of the most difficult hurdles that
a model must accomplish in the goal toward
building a particle physics model that explains
particle cosmology phenomena.  Thus, this fact
about the NMSSM has opened up the possibility
to build perhaps the minimal model that is consistent
with the Standard Model at low energy and particle
cosmology at high energy.
This is a current area of research that is being
explored.

\section{Testing warm inflation with observation}
\label{sect12}

In the last decade cosmology has transformed itself from being
a data starved to data rich field.
The primary data is of the CMB and the source of this data
is mainly through the two satellite experiments 
mentioned earlier COBE and WMAP.
These satellites basically have been collecting CMB photons
from all directions in the sky over the period of years.
From this, a map has been produced of the temperature
of the CMB at different locations on the sky,
parametrized by the polar angle $\theta$ and azimuthal angle
$\beta$. As already mentioned the CMB temperature is almost
the same in all direction at $T_0=2.725K$ with variations
at different points on the sky at the level of
one part in $10^5$.  The most important and detailed information
about the early universe
is contained in these small temperature fluctuations
$\delta T(\theta, \beta) = T(\theta, \beta) - T_0$.

For any theoretical model of the early universe, one can
calculate its predicted 
CMB temperature and accompanying fluctuation spectrum.
A key meeting point between theory and observation in early
universe cosmology is on comparing model predictions
against CMB data.
In particular, the largest scale fluctuations 
in the CMB could not have been produced by a causal
mechanism in the Standard Cosmology, but 
as discussed in Sect. \ref{sect8} inflation offers
one of the most compelling explanations for their creation.
Moreover each inflation model would predict its own
distinct spectrum.  Recall from Sect. \ref{sect8} that during
inflation, a given fluctuation mode of the inflaton field will
expand along with the universe.  Eventually once the wavelength of
this mode becomes bigger than the Hubble radius
during inflation, this mode
freezes after which no causal physics can alter its amplitude.
The earlier a mode freezes out during inflation, the bigger
the scale of the fluctuation to which it corresponds. 
In particular, the largest fluctuation scale observed in
the universe today corresponds to a fluctuation mode
that froze about 60 efolds before the end of inflation.
Each subsequent mode that freezes out corresponds to
fluctuations in our universe today at successively smaller and
smaller length scales.  Thus during inflation, each mode that freezes
in general will reflect the state of the inflaton field
at that time, and this will vary slightly from one
mode to the next as they freeze-out.  These small variations
will differ from model to model, and so each model
will show its own characteristic signature in its prediction for the
temperature fluctuations of the CMB. 
 
As it turns out, the observed fluctuations in $\delta T(\theta,\beta)$
agree very well with the case where the amplitude of
every mode that freezes during inflation is
exactly the same,   This is known as a scale invariant,
or Harrison-Zeldovich, spectrum.
The fact that the spectrum produced in scalar field slow roll
inflation is almost scale invariant was another of the successes
of inflation in the early days.
Today, what observers and theorists are looking for 
is how much the data deviates
from scale invariance, and which models of inflation
are consistent with these deviations.

Data has shown some interesting types of deviations from scale
invariance, although it is still early days in the field
for definitive claims.   
For example, the first year WMAP data indicated the interesting
feature that the amplitude of density fluctuations at the
largest scales are somewhat suppressed to those at the smallest
scales.  This is a discriminating feature which
could be used to rule out models.
In cold inflation, this feature turns out to be very
difficult to explain, although models do exist \cite{cirun}.
In contrast, for warm inflation this feature can be
explained in some common models, with the important point
being that dissipative effects play the key role \cite{hmb}.
Thus if future data supports this first year WMAP finding,
it may prove useful in distinguishing the two forms of
inflationary dynamics.

There are other interesting features
that can arise from dissipative effects.
One example is oscillations
in the CMB temperature fluctuation 
amplitude as one varies scales \cite{hmb}.
Also, the nonzero vacuum energy during inflation induces
production of gravity waves.  The amplitude of gravity waves
is proportional to the scale of the vacuum energy, which
typically differs in warm versus cold inflation models.
So this offers another measurement that could assist in
determining the form of dynamics during inflation \cite{tb}.
The distribution of the fluctuations is another area
of investigation.  The CMB fluctuations appear to be
very close to gaussian distributed, but the small
nongaussian deviations could provide information
that discriminates between warm and cold inflationary
dynamics \cite{sg}.

\section{Outlook}
\label{sect13}

Inflationary cosmology today is at a crossroad. It has reached
the stage where data plausibly argues for the
existence of an inflation phase occurring sometime
in the early universe.  What is unclear at the moment
is the level of precision to which inflation can be discriminated.
What is clear is no single test will be adequate on its own
to firmly establish the existence of inflation.
In fact data alone will not be enough. Guidance
from quantum field theory also will be necessary
to narrow the possible inflation models.  The interplay between
theory and observation perhaps eventually will 
even single out just one model.

Warm inflation has played an important role in realizing the
prominent role that quantum field theory must play.
In the early days where cold inflation was believed to
be the only dynamical realization of inflation, the importance
of quantum field theory in inflation models was very limited.
The inflaton evolution equation was thought of basically
as a classical equation with quantum field theory entering
only to the extent of the effective potential
and the quantum fluctuations of the inflaton.
The introduction of warm inflation has shown that
interactions during inflation play a much more significant
role than just this.
The role of dissipation and
fluctuation are now seen to be significant 
features of inflation which
in fact could even be the main story during inflation.


\end{document}